# Possibility of using Parallel beam CT in dental treatment by correction of movement errors with virtual Alignment Method

**Seokhwan Yoon[1], Seong-Kyun Kim[1]\*, Kyungtaek Jun[2]\***

[1]Department of Prosthodontics & Dental Research Institute, Seoul National University Dental Hospital, School of Dentistry, Seoul National University, Seoul, South Korea

[2]Department of Applied Mathematics and Statistics, Stony Brook University, New York, USA

\* Correspondence: ksy0617@snu.ac.kr (S.K.), ktfriends@gmail.com (K.J.)

## Abstract

The purpose of this study is to correct the movement errors that may occur because of the patients' movement during the CT scanning and thus, to obtain more precise CT images, additionally to investigate the possible use of Parallel Beam CT in the dental treatment. An extracted human tooth was scanned with PBCT and the possible movement errors that patients may cause during the CT scan were intentionally assigned. The maximum width of errors was designed to reach 100 μm. The trajectory of the fixed point was traced in the sinogram that includes the fixed point. The part deviated from the sinusoidal curve was modified using the Virtual Alignment Method. The quality of reconstructed image was remarkably degraded when the movement errors were given. A radiopaque part inside the tooth was used as a fixed point. The errors were corrected by shifting the fixed point to the center of sinogram and this relocation of fixed point did not cause any distortion of image or change of size. The movement errors were corrected with the fixed point inside the tooth. The more precise CT images obtained through this method are expected to be of extensive application in the area of dental treatment using CAD/CAM.

# Introduction

CT scanning is applied to a broad range of diagnosis and treatment in medical area[1,2]. CT used in dental treatment area brings an improved quality of treatment and broadens the treatment possibility[3,4]. Especially the development of Cone-Beam CT, CBCT, made CT more accessible[5,6]. Digital dentistry is another area that brings improvement in dental treatment[7,8]. Oral or model scanner digitalizes the three dimensional information of a real object and this, with 3 D printer, milling machine, and CAD/CAM made digital dentistry possible.

The CTs that are used in laboratories are usually the Parallel Beam CTs, the PBCTs, the precision already reached beyond micrometer level to the nanometer level. Study and development of CT and the software to process the data from the CT go on very vigorously[9,10]. However, the images through medical CT have not reached an expected level in spite of development of other elements in CT[11,12]. There can be many reasons; assumably, the main reason should be that the living human is the object[13,14]. Unlike the samples in the laboratory, humans continuously breathe, and the fact that the CT scans them makes them nervous and uneasy, and consequentially, makes them tremble. These unintended movements result in the deterioration of the quality of CT images. According to a study, during the 10 to 20 seconds of CBCT scanning time, patients make movement errors in width of about 2 to 10 millimeters[13]. In other words, without this movement error not corrected, it is simply not possible to get an image of which resolution is higher than 1 millimeter, even when scanned with a most precise CT machine.

In the Virtual Alignment Method, VAM[15], the fixed point, that has a distinctive radiopacity with the environment inside or outside of the sample, is used. The sinogram is a set of projection image scanned with PBCT, and when there is no error, every position of the space including the fixed point shows sinusoidal curve. However, when the sample moves, the error deviates from the sinusoidal curve. It is hard to know whether the error occurred or not in the complicated sinogram, but when there is a fixed point that is distinctive from the environment, we can not only distinguish the movement errors, but also correct them by making the fixed point to follow the sinusoidal trace[16,17].

To apply the merit of PBCT, now commonly used in laboratories, to medical area, the patients' movement errors have to be solved. If they can be solved with VAM, there is a high possibility to use PBCT in medical area. This study is built on this possibility. In this study, human teeth were selected to be the sample and be scanned by PBCT, and the patients' movement errors were assumed and applied to the projection images. Using the fixed point and the VAM, the errors will be modified and ultimately the possible use of PBCT in dental treatment will be suggested.

# Materials and Methods

Using extracted human tooth, PBCT scanning is carried out and the movement errors that patients possibly make in the clinic were applied to the obtained projection image. This project images' sinogram including errors does not graphically satisfy Helgason-Ludwig Consistency Conditions, HLCC.[18]. In this study, the error in the sinogram is corrected using the fixed point inside the tooth and through this, the reconstruction image without error is attained. Protocols were approved by the Institutional Review Board of School of Dentistry, Seoul National University, and all individuals gave informed consent (IRB No. S-D20200014)).

### Attaining the CT images
PBCT scanning was performed with the beamline 6C Bio Medical Imaging of the Pohang Light Source-II[19]. The photon source of this beamline is a multi-pole wiggler and the beamline projects a monochromatic X-ray beam between 10 keV and 55 keV. For this study, we used a double crystal monochromator based on silicon (220) reflection (DCM-V2; Vactron, Daegu, South Korea) to produce a 45-keV X-ray beam illumination. The specimen was put on the air-bearing rotation stage (ABRS-150MP-M-AS, Aerotech, Pittsburgh), that is 36 meter away from the light source, and 920 millimeter behind the specimen lied the X-ray microsope (Optique, Peter, Lentilly, France). The X-ray microscope has a 100-µm-thick CdWO4 scintillator facing (010) direction (Miracrys LLC, Nizhny Novgorod, Russia), an objective lens with a magnification factor of 2 (PLAPON 2X/NA 0.08, Olympus, Tokyo, Japan), and a scientific CMOS camera (Edge 5.5, PCO AG, Kelheim, Germany)). The field of view (FOV) was 7.8 mm by 6.6 mm with a pixel size of 3.05 µm. For CT scanning, projections were acquired every 0.5 degree of a complete 180 degree rotation, and the exposure time to the X-ray was 1.0 second for each projection. To modify the image, we took 5 more beam-profile images and 5 dark-field images. Before reconstructing the CT image, we reduced the size of projection into 1/16, which brought the effective pixel size to 12.2 µm. We used Octopus 8.7 software (XRE, Gent, Belgium), which supported filtered-back projection (FBP) reconstruction to complete the CT reconstruction.

We scanned the tooth sample using Green16® (Vatech, Hwaseong, Republic of Korea) to compare the images shoot with PBCT and CBCT. Projection images 100 sheets per second were obtained for 18 seconds with an X-ray source of 94 kVp, 7.8 mA. FOV is 50 mm X 50 mm and the size of pixel is 80 µm. The projection images were reconstructed with Ez3D Plus software (Vatch, Hwaseong, Republic of Korea).

## Assigning the errors

The movement errors, which can occur during the CT scanning, is made by the patients. It is thought that the patients move when they respire. The maximum width of patients' movement error is set 100 µm considering that various tools and methods are generally in use to suppress patients' movements during the scanning.

Assuming it takes 3 seconds for an adult to complete a respiratory cycle, a patient will take 6cycles and make up and down movements 6 times during the 18 seconds of dental CBCT scanning. In addition, uncontrolled tremble might occur due to the nervousness and this tremble would randomly happen without any periodic regularity. Therefore, in this study the vertical respiratory errors were assigned with 3 second-cycle, within the maximum width of 100 μm (25, 50, 75, 100 μm) and the random horizontal trembling errors were assigned within the maximum width of 100 μm (25, 50, 75, 100 μm). In Fig. 1, we showed the comparative location of each case and the expected comparative location of the tooth in the projection image when the error occurs. When the movement error occurs to the same direction of the X-ray, there is not a change in the projection image (Fig. 1B), and when the movement error occurs to the vertical direction to the X-ray projection, the location of tooth changes according to the moved location (Fig. 1C,D).

**Correcting the errors**

In this study, the patients' movement errors were corrected using VAM and the fixed point. The raidopaque filling material, which remains inside of the canal after the root canal treatment, was used to designate the fixed point. (Fig. 2C) There can be only one fixed point or several.

First, the fixed point in each projection image was identified, and then the axial location is moved so that the specific fixed point can be located on the same axial plane. The sinogram, that will be used to correct the errors, is made from the specific axial plane of collected fixed points. If the sinogram has no error, its graphic pattern will satisfy the HLCC's pattern. However, if it includes errors, the trajectory of fixed point shows deviations from the sinusoidal curve; the location of image to the projection angle will be shifted in order to follow sinusoidal curve.

If the sinogram, that is a collection of projection images of rotating object, has no errors, the shift route toward every location will satisfy the following orbital function.

$$T_{r,\varphi}(\theta) = r * \cos(\theta - \varphi),\ 0° \leq \theta < 180°$$

Here, r is the distance from the center of rotation to the specific location. θ is the angle of X-ray projection, φ is the angle that forms between the projected X-ray and the extended line of fixed point from the center of rotation. The fixed point should be shown on the sinogram satisfying this orbital function. However, if the error is included and the fixed point shows deviations from the orbital function, the method of correcting error is to shift this to the orbital function.

Nonetheless, when we do not know the original orbital function of the scattered fixed point, it requires a lot of time and effort to correct each fixed point so that they can be located on the original orbital functions. Therefore, in this study we simplified this issue by assuming that the fixed point rotates on the center of rotation, not by modifying the fixed point to follow the original orbital function. The way is to modify the fixed point to locate on the center of sinogram. In the sinogram, the orbital function of center of rotation is when r=0, so it gets $T_{0,\varphi}(\theta)$, and it turns up as a straight center line in sinogram. And then the locations will be modified so that the r value of fixed point to each θ will be 0. In this way, the error regarding the center of rotation can be corrected.

## Results

The tooth used in this study was the one with completed root canal treatment and there remained radiopaque filling material in the canal, which was used as a fixed point in the VAM. Fig. 2A is one of the projection images. The part marked with red dotted line is $273^{th}$ pixel location on the axial direction and it is the part used to construct reconstructed image and the sinogram in this study. In Fig 2C, the part marked with an arrow shows what was used as the fixed point.

In the images scanned with PBCT, we can clearly identify not only the tooth shape and crack line but also the radiopaque materials inside the canal (Fig. 2C). On the contrary, when scanned with CBCT, the images showed severe diffusion due to the raidopaque material inside the canal (Fig. 2B). The appearance of the tooth or canal shape, as well as the existence of crack line, is not identifiable.

Fig. 3A is the sinogram of the data from the $273^{th}$ pixel location toward the axial direction of the projection image. The whole shape of singoram including the fixed point seems to follow the sinusoidal curve successively. A Sinogram alone is not enough to know if there is an error. However, if we reconstruct the center of rotation based on the sinusoidal curve the trajectory of fixed point draws, the center of rotation lies on the location 1 pixel off from the center of the sensor. This is an error occurred since the center of sensor and the center of rotation of sample's rotation plate do not align.

## The result of patients' movement error

Fig. 4 shows the result of movement errors when the patients make during the CT scan. The width of vertical and horizontal movement was set 25, 50, 75, 100 µm, and the sinogram and reconstructed image of each case were shown. We can confirm that the trajectory of fixed point gets vague when an error occurs. When the error of 100 µm width occurs, the crack line is not distinguishable both in the sinogram and the reconstructed image.

## The result of correcting errors

The rotation axis should lie on the 320th pixel if there is no error in the 640 pixel-CCD and the center of rotation and the center of X-ray sensor align. However, the rotation axis calculated with the fixed point lies on 321th pixel. Here we added 2 more pixels to the lower part of the sonogram in order to modify the rotation axis error. In this way, the number of the pixels in the sinogram will be 642 and the center will be 321th. By doing this, we were able to align the rotation axis to the center of the sonogram. Fig. 3 shows the sinogram and the reconstructed image obtained after correcting the error.

Fig. 5A is a modified sinogram of Fig. 3, where the total sinogram is placed on the center of the broader virtual space. It is the concept that we extend the space vertically toward rotation axis, which is the center of CT scanning. Some pixels were added on the upper and lower part of the sinogram, so that the total will be 1240, and the rotation axis that the orbital function of the fixed point makes is shifted on the rotation axis of the extended sinogram. The sinogram was shifted as a whole and the inner trajectories were remained the same. The orbital function of fixed point, that was obtained by doing this, is $T_{191,9°} = 191\cos(\theta - 9°)$. In other words, the fixed point in the virtual space rotates on the 191th pixel apart from the center of rotation.

If we horizontally shift the tooth in the extended virtual space and locate the fixed point on the rotation axis, the fixed point during the CT scan will remain on the rotation axis without any change of location and the orbital function will be $T_{0,\varphi} = 0$. Because the fixed point is on the rotation axis, the trajectory will be presented as a straight line on the center of sinogram as shown in Fig. 5B. Although the shape of whole sinogram gets different from the one before shifting the fixed point, each point flows well with the sinusoidal curve. Fig. 6A is the reconstructed image using the Fig. 5B's sinogram in which we shifted the fixed point to the center of rotation. The difference between those two images is shown in Fig. 6B, which overall shows half a pixel margin of error.

## Discussion

In this study, we try to exam the possible application of PBCT in the dental treatment area by correcting the errors that can occur from the patents' movement during the PBCT scanning, through VAM using the human tooth sample. In VAM, the fixed point identifiable in the sinogram is used, and in this study, we designated the radiopaque part inside the tooth sample as the fixed point shown as an arrow in Fig. 2C. Just like this, anything of which trajectory is traceable in the sinogram can be used as a fixed point. If there is not any part inside can be used as a fixed point, any raidiopaque material like light-cured resin attached outside the tooth can be CT-scanned and used as a fixed point.

Any part that is inside of the rotating space appears in the sinogram satisfying the orbital function of $T_{r,\varphi}(\theta)$ if the images projected from the angles with regular interval were collected. The there is no error, the center of sinogram will be the center of rotation and its orbital function will be a straight line satisfying r=0, $T_{0,\varphi}(\theta)$. What we called a fixed point should appear in the sinogram and follow the orbital function. So if the fixed point does not draw the sinusoidal function in the sinogram, it should be modified to follow it, and by doing so, the errors will be corrected. The amplitude and the starting point of orbital function will vary depending on the location of fixed point. If it is far from the center of rotation, the amplitude will be wider and vice versa. If the fixed point is on the rotation axis, the amplitude of the fixed point will be 0 and located on the center in the sinogram. How to shift the fixed point to the rotation axis is to move the images from each projected angle in the sinogram, so that the fixed point is located on the center of the sinogram. We do not need complicated calculations or formulas, if we know the distance between the center and the fixed point on the projection image, we can always shift the fixed point to the center of rotation.

In order to shift the fixed point to the center of rotation, we need to get out from the pixel area obtained from the CT scanning. If we try to correct the rotation axis error of the sinogram in the 640 pixels of Fig. 3, the data from certain part should be cut off. Furthermore, if the fixed point is shifted to the rotation axis, the whole shape of the sinogram will change from the original straight line to the sinusoidal curve of which the center is the fixed point. (Fig. 5B) In this case, if we insist to look at this within only 640 existing pixels' frame, we will lose a lot of information. When the changes like in Fig. 5B is to be modified in 640 pixels, the large area of upper and lower part of the sinogram should be cut off as seen in the Fig. 7. If we use these data to reconstruct images, they would not provide us with correct information.

VAM solved this problem in the virtual space. It does not try to find the real rotation axis or error in the given information area, but to extend the space itself, so that the fixed point lies on the center in virtual space. That is to say, the real rotation does not lie on the center of the sinogram, but the amplitude of a sinogram is changed so that the fixed point we choose lies on the center. This kind of change does not affect the quality or size of the original reconstructed image. Fig. 6B is an overlapped image of Fig. 3B, which is a reconstructed image with rotation axis error alone modified and Fig. 6A, which is a reconstructed image in the virtual space of which the center is the fixed point. As we can see in the figure, there is no size difference or distortion of overall shape, the only observed difference is half-a-pixel density difference. This difference is inevitable in any modifying digital images.

The ultimate purposes of this study is to correct the patients' movement errors during the CT scan without a retake, and to obtain the clearest image possible in the given environment. Nevertheless, as shown in Fig. 4, if the patient's movement error is 100 μm, it is hard to identify the shape or the inner structure of the sample even with the most precise machine. This might be why PBCT has not been actively applied to medical area. There is no point of using PBCT in medical area if the quality of image is the same with cheap and convenient CBCT. However, if we can push the mechanic limit of the image quality of PBCT up to the best possible result, the future direction of researches on PBCT will be different from the ones in the past. This expected accumulation of researches will push forward the development of PBCT in medical area.

If we can utilize the precise CT image without errors in dental area, it will be possible to make a more accurate diagnosis and further, to apply to a direct treatment area like making dental prosthetic appliances. Recent developments of oral scanner, CAD/CAM, and dental materials already made digital dental treatment possible[20-22], and the range of treatment has been broadened. Only with three-dimensional information about abutment and adjacent teeth, it is possible to produce prosthetic appliances without a model tooth. When using the oral scanner, which is a tool to get information about image, the precision is certainly the issue of research, but whether it is possible to scan every part of the abutment tooth is also another important issue. For the oral scanner is directly used in human mouth, the issues like the anatomical limit of patients' mouth, the level of practitioner's proficiency, the contamination of abutment tooth due to saliva or blood, and the boundary of abutment tooth that lies on the lower part of gingival, hinder us from getting correct information. If we have precise and credible CT information about the teeth, obtained outside the human mouth, we will be able to get clear image of abutment teeth boundaries and it will certainly supplement the lacking information from the oral scanner. Furthermore, it is expected that we will be able to directly produce prosthetic appliances with the information from the CT scanning.

## Conclusion

Unlike the precise CTs used in laboratories, medical CTs have a huge limit in that the sample is the living human beings. The movements they make during the CT scanning prevent us from getting precise images. The errors, which occurred because of this problem, were corrected simply using VAM and fixed point. The reconstructed image through this correction process was almost the same with the one without the error in terms of size and the level of precision; we were able to get the same quality of image with the original image. Within the limits of this study of which the focus is on the tooth, high quality of projection image was obtained by correcting the movement errors. If we can use this precise

CT images in producing prosthetic appliances using CAD/CAM, PBCT is expected successfully adopted in the dental treatment area.

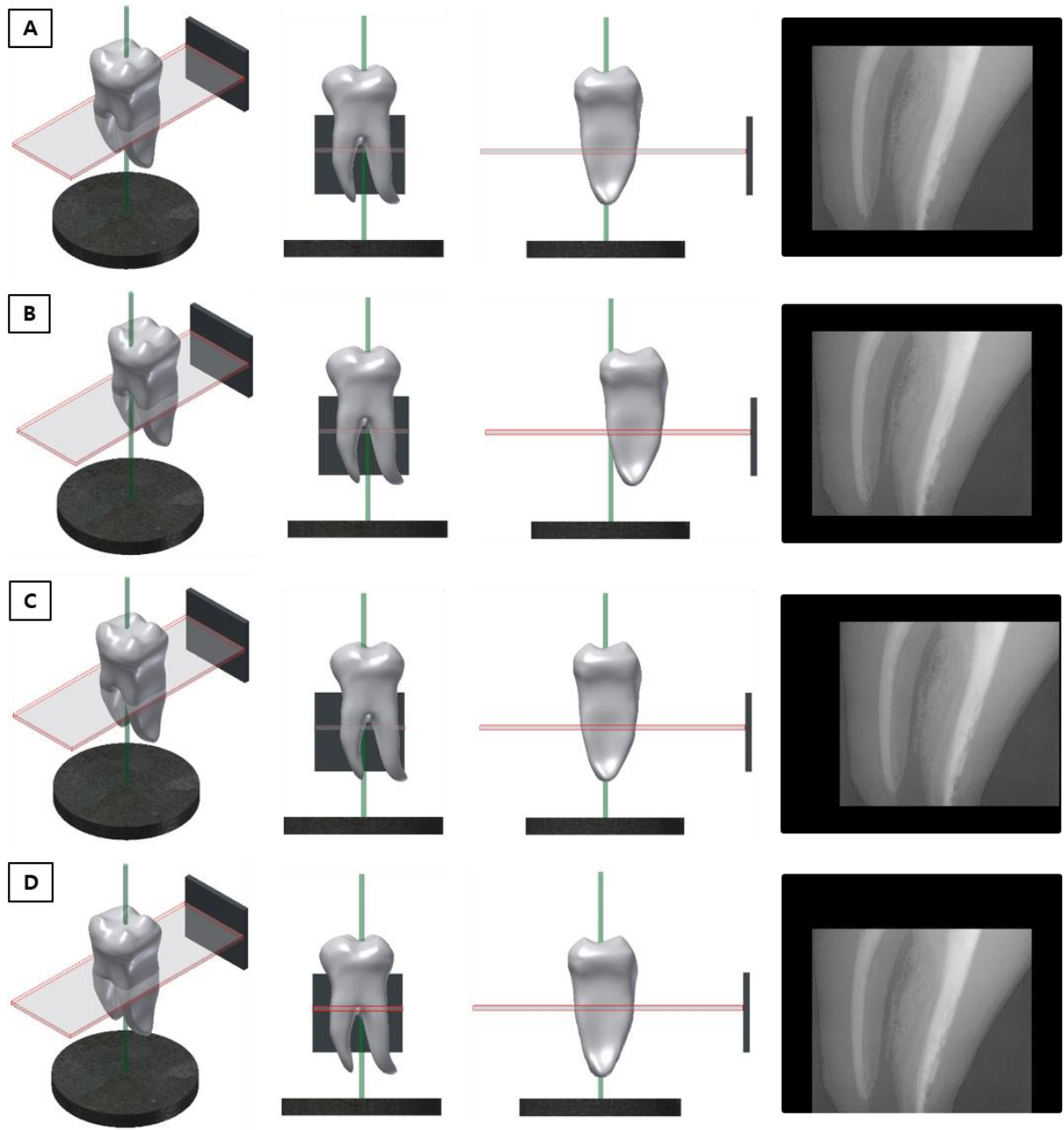

**Fig. 1. The diagrams and the projection images of the types of movement errors that can happen during the CT scanning** (A) when there is no error, (B) when the movement error occurred to the direction of X-ray projection, (C) when the movement error occurred to the horizontal direction of X-ray projection, (D) when the movement error occurred to the vertical direction of X-ray projection.

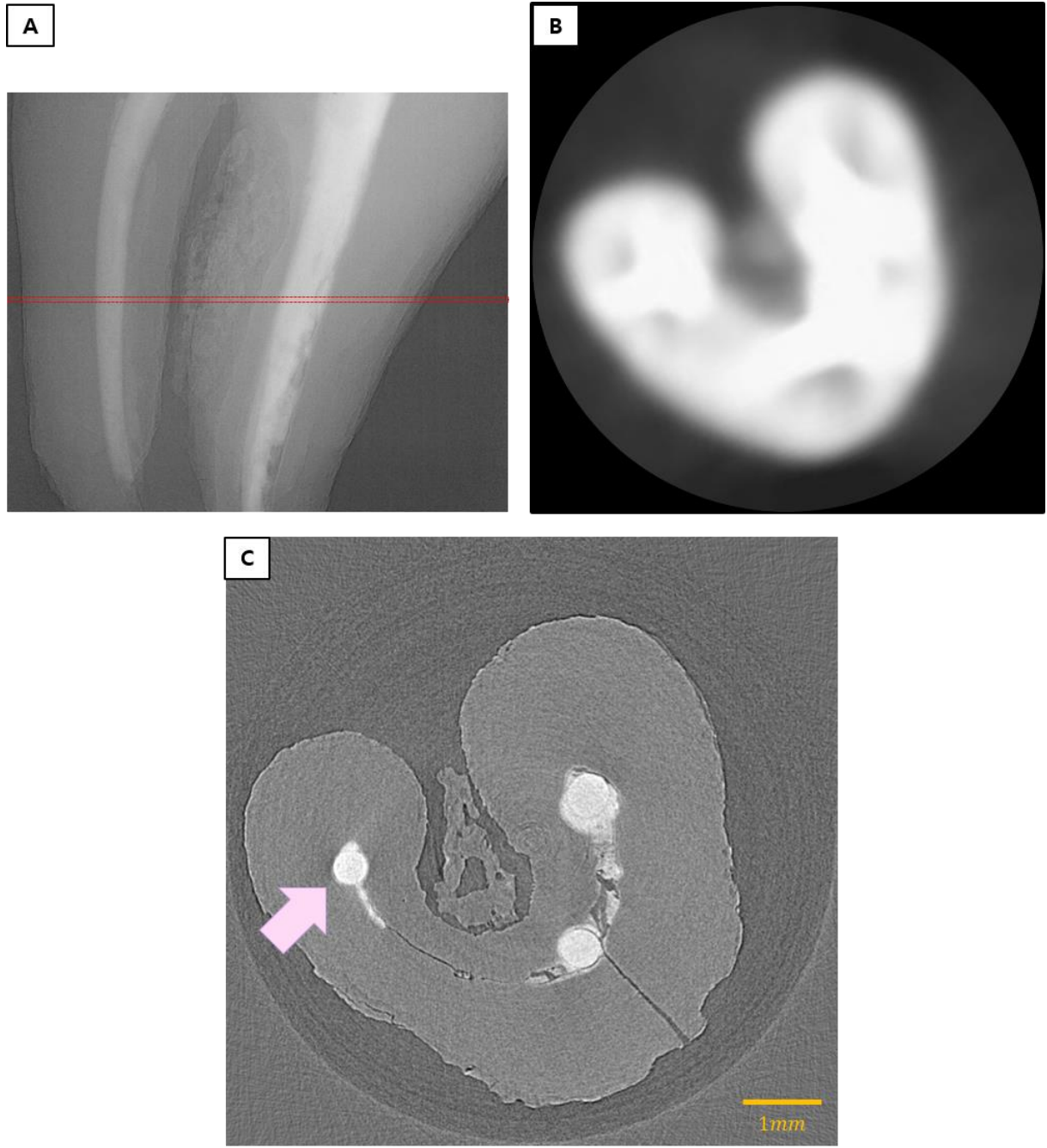

**Fig. 2. The image of tooth that was scanned with PBCT and CBCT** (A) the projection image of tooth with PBCT. The red dotted line shows 273th location to the axial direction and it is the standard location of the sinogram and the reconstructed image. (B) The reconstructed image of tooth scanned with CBCT. (C) The reconstructed image of tooth scanned with PBCT. The arrow shows the part used as the fixed point.

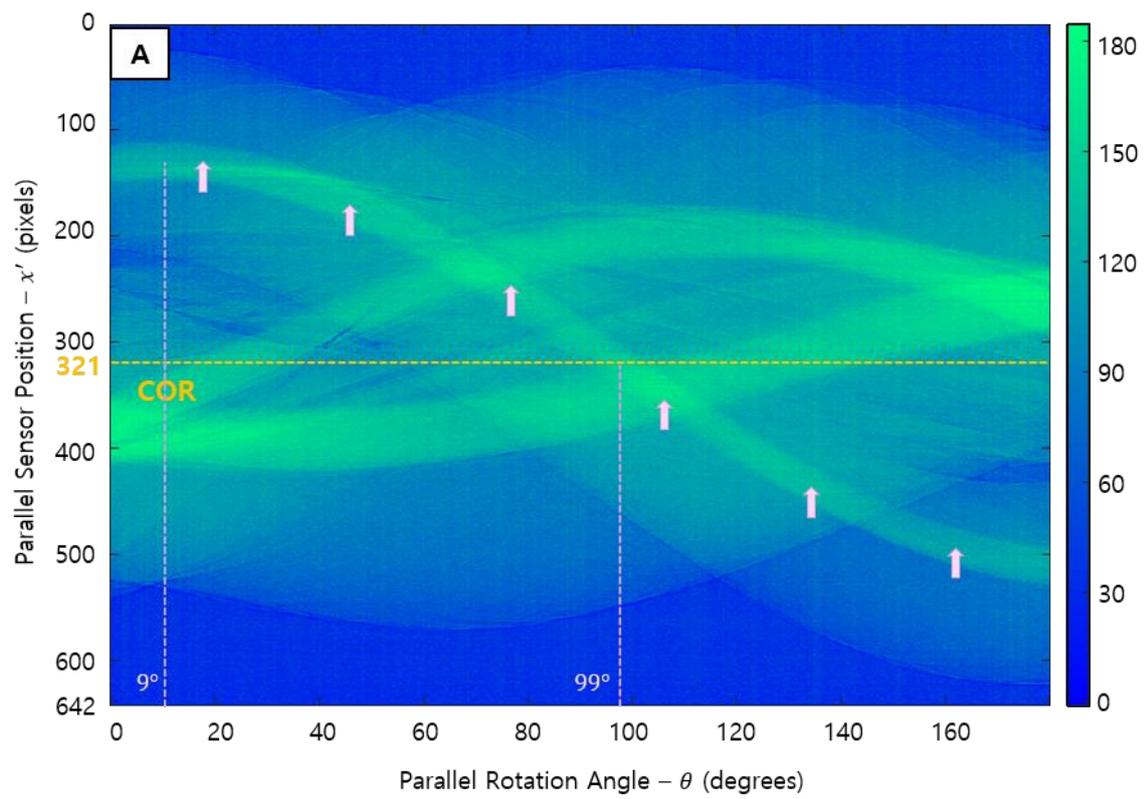

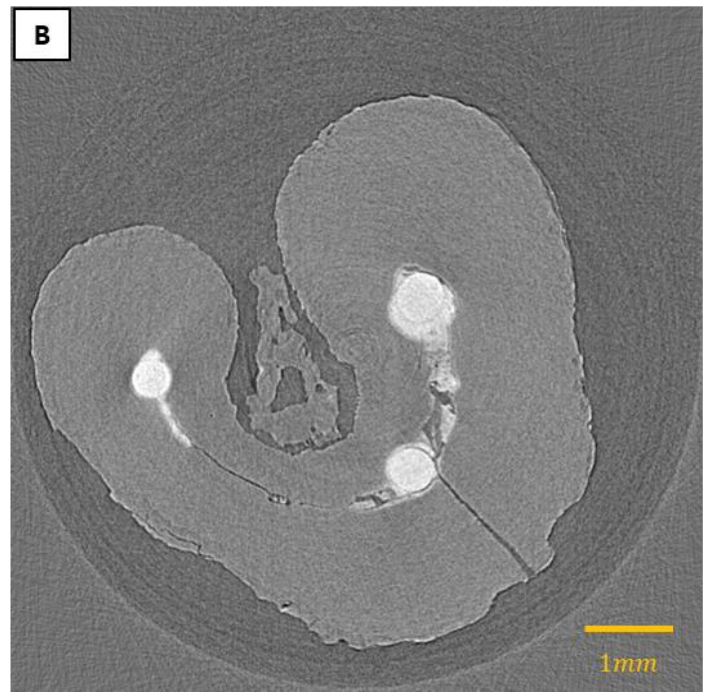

**Fig. 3. The sinogram and reconstruction image after correcting rotation axis offset.** (A) The sinogram of 273th pixel location toward the axial direction of projection image when scanned with PBCT. The arrows designate the trajectory of fixed point. (B) The reconstruction image.

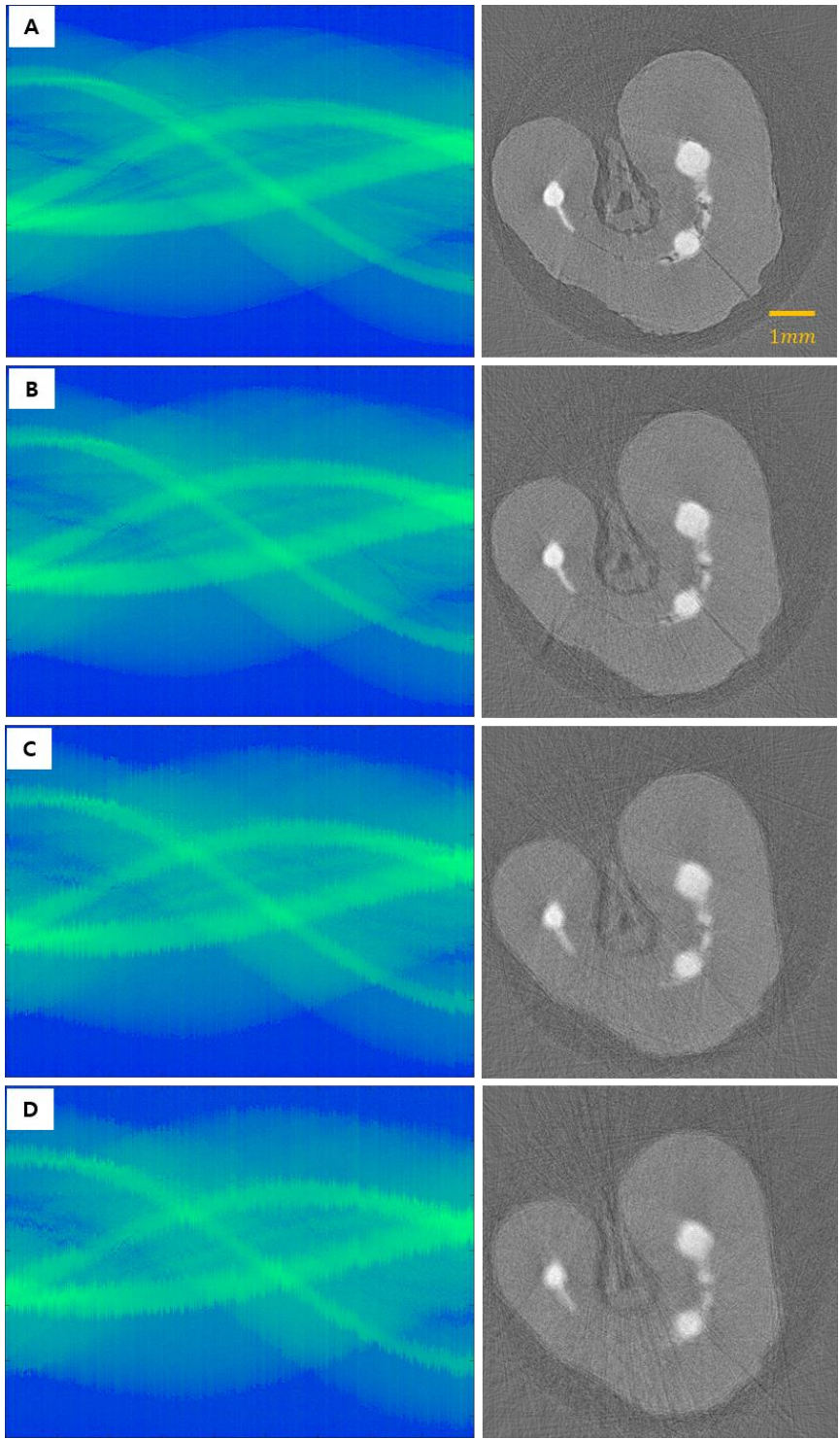

**Fig. 4. The sinogram and the reconstructed image when the vertical vibration movement and the random horizontal movement were assigned.** Each shows the errors width of (A) 25 µm, (B) 50 µm, (C) 75 µm, (D) 100 µm

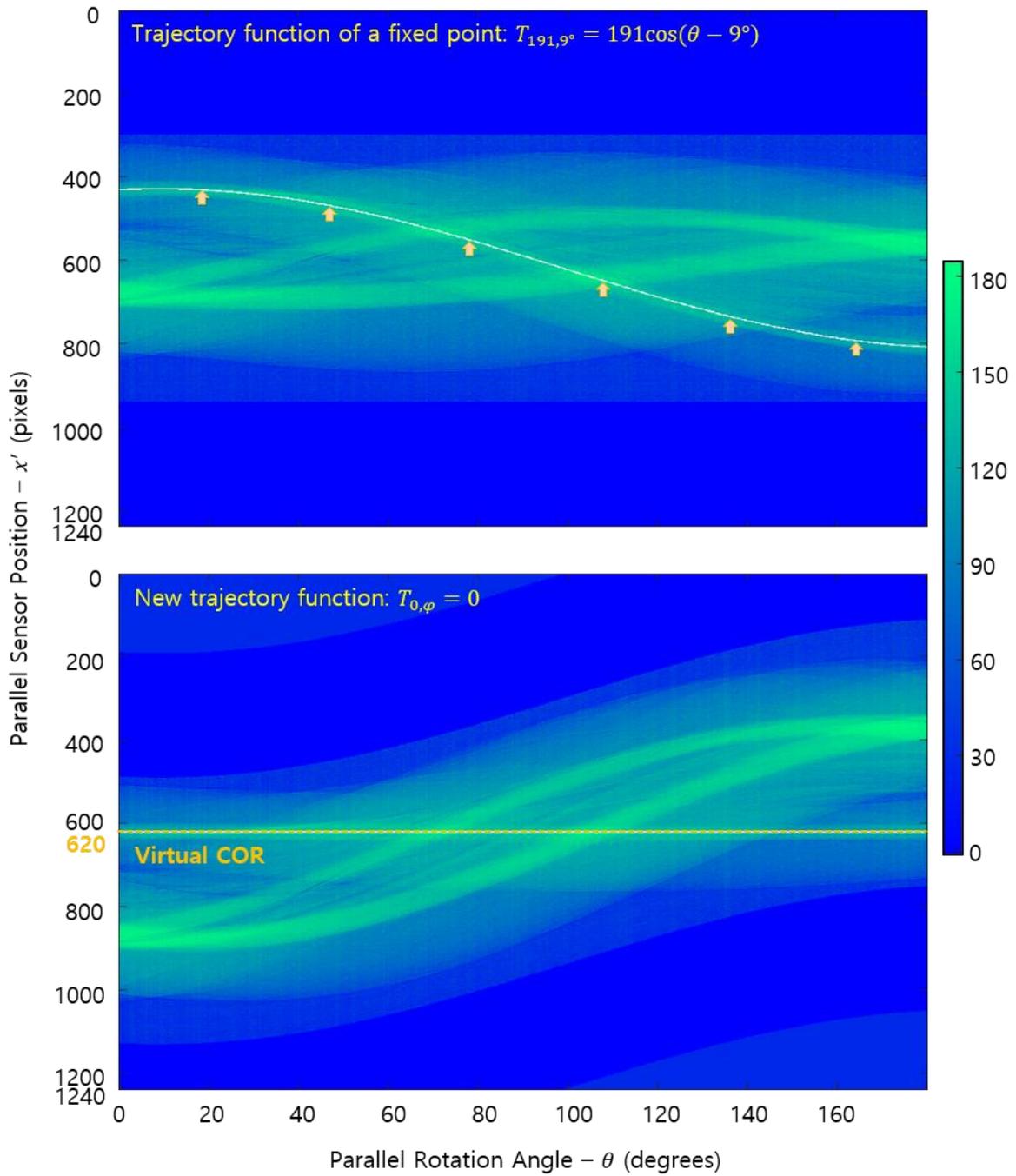

**Fig. 5. The sinogram extended to the virtual space.** (A) The sinogram where the fixed point follows the orbital function of $T_{191,9°} = 191\cos(\theta - 9°)$. (B) The sinogram in which the fixed point is shifted to the rotation axis and the orbital function is changed into $T_{0,\varphi} = 0$.

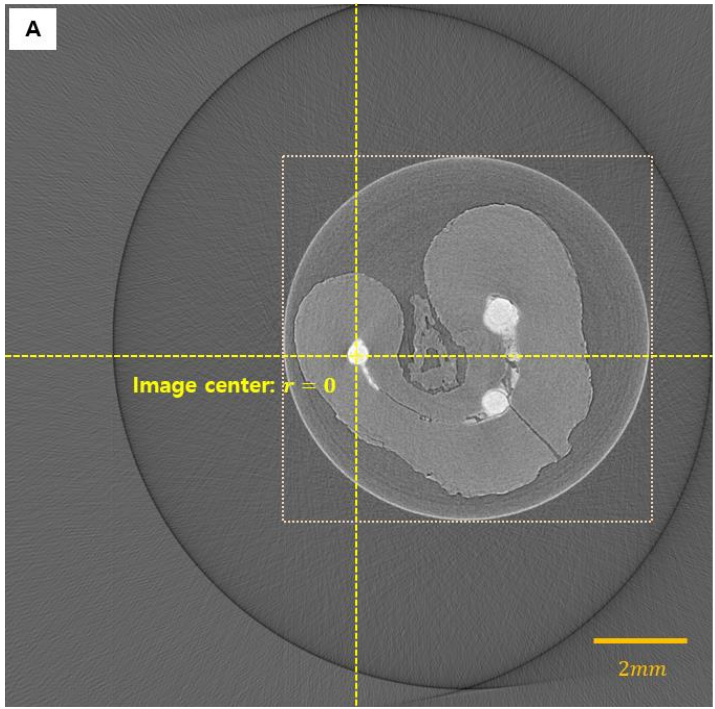
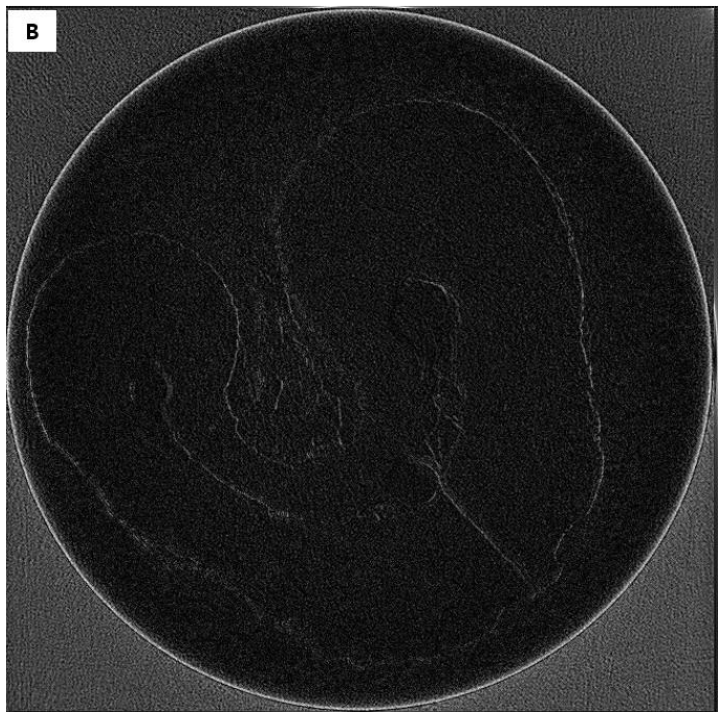

**Fig. 6. The reconstructed image after correcting errors** (A) The reconstructed image after the fixed point is shifted to the rotation axis in the virtual space. (B) The overlapped margin of the images before and after VAM.

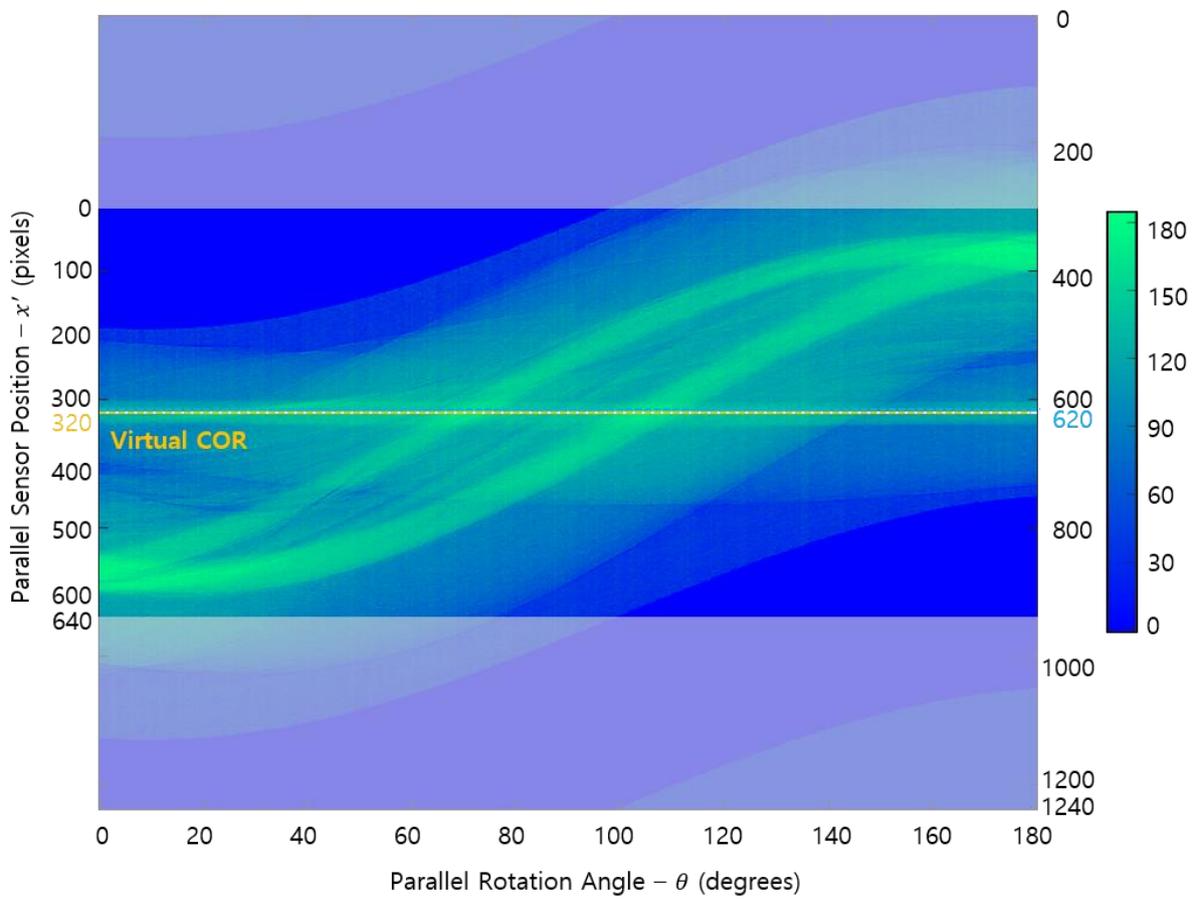

**Fig. 7. The sinogram in which the fixed point is shifted to the center of rotation.** The original 640 pixels are maintained and the gray part of the picture signifies the part outside of 640 pixels area that is not reflected in the reconstructed image because of this change.